# Gene Expression Data Knowledge Discovery using Global and Local Clustering

Swathi. H

**Abstract**—To understand complex biological systems, the research community has produced huge corpus of gene expression data. A large number of clustering approaches have been proposed for the analysis of gene expression data. However, extracting important biological knowledge is still harder. To address this task, clustering techniques are used. In this paper, hybrid Hierarchical k-Means algorithm is used for clustering and biclustering gene expression data is used. To discover both local and global clustering structure biclustering and clustering algorithms are utilized. A validation technique, Figure of Merit is used to determine the quality of clustering results. Appropriate knowledge is mined from the clusters by embedding a BLAST similarity search program into the clustering and biclustering process. To discover both local and global clustering structure biclustering and clustering algorithms are utilized. To determine the quality of clustering results, a validation technique, Figure of Merit is used. Appropriate knowledge is mined from the clusters by embedding a BLAST similarity search program into the clustering and biclustering process.

―――――――― ◆ ――――――――

## 1 INTRODUCTION

THE clustering is the process of grouping data into classes or groups so that objects within a cluster have high similarity in comparison to one another, but are very dissimilar to objects in other classes [11]. Clustering can also facilitate taxonomy formation,that is,the organization of observations into a hierarchy of classes that group similar events together.There exist a large number of clustering algorithms in the literature.The clustering algorithms are commonly applied in molecular biology for gene expression data analysis [5, 6]. These algorithms are used to partition genes into groups based on the similarity among their expression profiles. These clustering algorithms can be broadly classified into partitional and hierarchical algorithms [11].

   The partitional clustering algorithms generate a single partition, with a specified or estimated number of nonoverlapping clusters, of the data in an attempt to recover natural groups present in the data [11]. Hierarchical clustering (HC) algorithms construct a hierarchy of partitions, represented as a dendogram in which each partition is nested within the partition at the next level in the hierarchy [11]. The most commonly used partitional clustering algorithms are K-Means (KM) and k-mediods [11]. The KM algorithm takes the input parameter k, and partitions a set of n objects into k clusters so that the resulting clusters have high intracluster similarity and low inter cluster similarity. Cluster similarity is measured as the mean value of the objects in a cluster, which can be viewed as the cluster's centre of gravity [11].

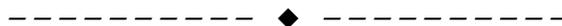

- H.Swathi is with the Department of Information Technology, Vivekanandha College of Engineering for Women, Tiruchengode, India.

However both KM and HC clustering algorithm have certain disadvantages like difficulties in specifying the number of clusters in advance and in selection of merge or split points [11]. HC cannot represent distinct clusters with similar expression patterns. As clusters grow in size, the actual expression patterns become less relevant [11]. KM clustering requires a specified number of clusters in advance and chooses initial centroids randomly; in addition, it is sensitive to outliers [11]. A novel hybrid approach that combines the merits of these two methods and discards their innate disadvantages [1]. HC is carried out first to decide the location and number of clusters in the first round and run the KM clustering in next round. This approach provides a mechanism to handle outliers [1], [2], [3], [12].

When clustering data the similar observations should be grouped together. Thus needs to be able to compute the distance between two data objects, but it can be defined in many forms [12].Distance measurements influence the shape of the clusters, as some elements may be close to one another according to one distance and farther away according to another[16]. In this paper the Pearson's Correlation Coefficient measurement is used to calculate the distance.In this work the gene expression data is clustered by global and local clustering.

Gene expression is the process by which inheritable information from a gene, such as the DNA sequence,is made into a functional gene product,such as protein or RNA[15].The expression of many genes is regulated after transcription(i.e., by microRNAs or ubiquitin ligases) and an increase in mRNA concentration need not always increase expression.The advances in microarray technology,high-throughput and low-throughput methods such as "tag based" technologies like Serial Analysis of Gene

Expression(SAGE) or the more advanced version Super-SAGE,which can provide a relative measure of the cellular concentration of different messenger RNAs[14].The expression levels of large numbers of genes in a tissue at different time points and also the relative amounts of mRNA produced at these time points provide a gene expression time series for each gene. The time series gene expression data consists of a matrix containing intensity data for a group of genes for certain time points [17].

The process of evaluating the results of a clustering algorithm is called cluster validity assessment.Two measurement criteria have been proposed for evaluating and selecting an optimal clustering scheme [4]:

Compactness: The member of each cluster should be as close to each other as possible.A common measure of compactness is the variance.

Separation: The clusters themselves should be widely separated.

The cluster validation procedures divided into two main categories:

External criterion analysis
Internal criterion analysis

External criterion analysis validates a clustering result by comparing it to a given "gold standard"which is another partition of the objects[18].The gold standard must be obtained by an independent process based on information other than the given data set.There are many statistical measures that assess the agreement between an external criterion and a clustering result.

For validation of clustering results,external criterion analysis has the strong benefit of providing an independent , hopefully unbiased assessment of cluster quality.On the other hand,external criterion analysis has the strong disadvantage that an external gold standard is rarely available.Internal criterion analysis avoids the need for such a standard,but has the alternative problem that custers are derived[19]. Different clustering algorithms optimize different objective functions or criteria.Assessing the goodness of fit between the input data set and the resulting clusters is equivalent to evaluating the clusters under a different objective function.

## 2 BICLUSTERING GENE EXPRESSION DATA

### 2.1 Pearson's Correlation Coefficient

The Pearson's correlation coefficient, which measures the similarity between the shapes of two expression patterns (profiles) [12]. Given two data objects and Pearson's correlation coefficient is defined as

$$\text{Pearson}(o_i, o_j) = \frac{\sum_{d=1}^{P}(o_{id} - \mu_{oi})(o_{jd} - \mu_{oj})}{\sqrt{\sum_{d=1}^{D}(o_{id} - \mu_{oi})^2}\sqrt{\sum_{d=1}^{P}(o_{jd} - \mu_{oj})^2}} \quad , \quad (1)$$

where $\mu_{oi}$ and $\mu_{oj}$ are the means for $\vec{o_i}$ and $\vec{o_j}$ respectively. Pearson's correlation coefficient views each object as a random variable with observations and measures the similarity between two objects by calculating the linear relationship between the distributions of the two corresponding random variables [12]. Pearson's correlation coefficient is widely used and has proven effective as a similarity measure for gene expression data. It is not robust with respect to outliers, thus potentially yielding false positives which assign a high similarity score to a pair of dissimilar patterns [12].

### 2.2 Figure of Merit

The Figure of Merit (FOM) methodology is used for assessing the quality of clustering results. FOM is a scalar quantity, which is an estimate of the predictive power of a clustering algorithm [4]. A typical gene expression data set contains measurements of expression levels of n genes under B conditions. Assume that a clustering algorithm is applied to the data from condition 1, 2, 3... (e-1), (e+1)... B. The condition *e* is used to estimate the predictive power of the algorithm [4]. The FOM under the condition e is defined

$$\text{FOM}(e,k) = \sqrt{\frac{1}{n} * \sum_{i=1}^{k} \sum_{x \in C} (R(x,e) - \mu_{ci}(e))^2} \quad , \quad (2)$$

where R (g, e) is the expression level of gene g under condition e, $\mu_{ci}$ (e) is the average expression level in condition e of genes in cluster $C_i$.

### 2.3 Biclustering

The clustering has proved to be a powerful tool for data analysis and continues to be an active area of research [5]. However when applied to microarray data, clustering techniques have certain difficulties. The problem derives from the fact that when analyzing a microarray data matrix, conventional clustering techniques allow one to cluster genes (rows) and thus compare expression profiles, or to cluster conditions (columns) and thus compare experimental samples but are not intended to allow one to accomplish both simultaneously [5]. This biclustering approach is capable of discovering local patterns in microarray data. Each gene in a bicluster is selected using only a subset of the conditions and each condition in a bicluster is selected using only a subset of the genes [5]. There are many biclustering algorithms aimed at discovering biclusters.

### 2.4 Bioinformatics Tools

Using bioinformatics tools, the differential gene expression can be studied, which could lead to the identification of important gene/proteins (which were not reported previously) and invasion mechanism. For obtaining the sequence information about the genes the sequence similarity search tools are used. Therefore, the bioinformatics tool **BLAST** is utilized to identify sequences similar to the query sequences.

### 2.5 Basic Local Alignment Search Tool (BLAST)

Basic Local Alignment Search Tool, or BLAST, is a program for comparing primary biological sequence infor-

mation, such as the amino-acid sequences of different proteins or the nucleotides of DNA sequences [9]. A BLAST search enables a researcher to compare a query sequence with a library or database of sequences, and identify library sequences that resemble the query sequence above a certain threshold. It is about 50 times faster than dynamic programming [9].

Dynamic programming is a method of solving complex problems by breaking them down into simpler steps. It is applicable to problems that exhibit the properties of overlapping subproblems and optimal substructure. **BLAST** seeks out local alignment (the alignment of some portion of two sequences) as opposed to global alignment (the alignment of two sequences over their entire length). By searching for local alignments, **BLAST** is able to identify regions of similarity with two sequences.

## 3 RELATED WORK

### 3.1 Statistical-Algorithmic Method for Bicluster Analysis (SAMBA)

Tanay et al., [6] introduced Statistical-Algorithmic Method for Bicluster Analysis (SAMBA). It is a graph-theoretic approach, in combination with a statistical data modeling. In SAMBA framework, expression matrix is modeled as a bipartite graph. A bicluster is defined as a subgraph, and a likelihood score is used in order to assess the significance of observed subgraphs [8]. This algorithm is applied to yeast and human clinical data. The clusters obtained are superior to the Cheng and Church biclustering approach. Moreover, the results differentiating fine tissue types from DLBCL (infected) tissues.

### 3.2 Biclustering by Iterative Genetic Algorithm (BIGA)

The Biclustering by Iterative Genetic Algorithm (BIGA) [7] approach is proposed to identify transcriptional module (TM) in gene expression data, avoiding the intrinsic limitations of the heuristic biclustering algorithms. Every TM is composed of the gene subset and the condition subset from the original gene expression data and also possesses *a* significant level of correlativity requested by a user. Besides, a novel fitness function for a statistically significant and condition-specific cluster, i.e., the *a*-TM, is defined [9].

### 3.3 Flexible Overlapped Biclustering (FLOC)

Flexible Overlapped Biclustering (FLOC) introduced by Yang et al., [8]. It starts from a set of seeds (initial biclusters) and carries out an iterative process to improve the overall quality of the biclustering. After each iteration, each row and column is moved among biclusters to produce a better bi-clustering in terms of lower mean squared residues. The best biclustering sub matrix obtained will serve as the initial biclustering for the next iteration. The algorithm terminates when the current iteration fails to improve the overall biclustering quality [8].

### 3.4 Robust Biclustering Algorithm (ROBA)

A Robust Biclustering Algorithm (ROBA) [10] is a simpler one because; it uses basic linear algebra and arithmetic tools. ROBA is made up of three parts. The first part consists of performing the data conditioning, to get rid of the noise and to solve the problem of missing values. The second part consists of decomposing the data matrix A into its elementary matrices, and the last part is used to extract user defined biclusters [10].

### 3.5 xMOTIFs (biclusters)

Murali and Kasif [13] assumed that data may contain several xMOTIFs(biclusters) and aimed at finding the largest xMOTIF: the bicluster that contains the maximum number of conserved rows.The merit function used to evaluate the quality of a given bicluster is thus the size of the subset of rows that belong to it.Together with this conservation condition,an xMOTIF must also satisfy size and maximal properties: the number of columns must be in at least a α-fraction of all the columns in the data matrix,and for every row not belonging to the xMOTIF the row must be conserved only in a β-fraction of the columns in it. Ben-Dor et al. considered that row (genes) has only two states (up-regulated and down-regulated) and looked for a group of rows whose states induce some linear order across a subset of the columns (conditions). This means that the expression level of the genes in the bicluster increased or decreased from condition to condition. Murali and Kasif [13] consider that rows (genes) can have a given number of states and look for a group of columns (conditions) within which a subset of the rows is in the same state.

### 3.5 Order-Preserving Sub-Matrix (OPSM)

Ben-Dor et al.defined a bicluster as an order-preserving sub-matrix (OPSM)[13].According to their definition,a bicluster is a group of rows whose values induce a linear Order across a subset of the columns.This work focused on the relative order of the columns in the bicluster rather than on the uniformity of the actual values in the data matrix.More specifically, they want to identify large OPSMs.A sub-matrix is order preserving if there is a permutation of its columns under which the sequence of values in every row is strictly increasing.

Although the straightforward approach to the OPSM problem would be to find a maximum support complete model, that is, a set of columns with a linear order supported by a maximum number of rows, Ben-Dor et al. aimed at finding a complete model with highest statistically significant support.The statistical significance of a given OPSM is thus computed using an upper-bound on the probability that a random data matrix of size n-by-m will contain a complete model of size s with k or more rows supporting it.In the case of gene expression data such a sub-matrix is determined by a subset of genes and a subset of conditions, such that, within the set of conditions, have the same linear ordering.

## 4 RESEARCH METHODOLOGY

The hybrid clustering combines both HC and KM. In this method, an agglomerative HC is carried out first. In this bottom-up strategy to start with each objects in its own cluster and then merges these atomic clusters into larger clusters, until all the objects in the single cluster. The shortest pairwise distance between elements of the two clusters is used in pairwise single-linkage clustering. At the end of pairwise single-linkage cluster a tree will be formed. Afterward, the within-cluster distance between any points in the cluster is computed in Pearson's correlation coefficient.

Based on the distance, cluster formation is done by Hashing function. The bucket table is referred as clusters and it indexed as 0, 1…N-1. In the hash function the buckets will be roughly equal in size, so the list for each bucket will be short. If there are M genes in the set, then the average bucket will have N/M genes. N will estimated and choose M to be roughly as large, then the average bucket will have only one or two genes. It will decide the K-value for the KM. The initial run of the KM is also taken from the HC. This hybrid algorithm automatically finds good initial centroids for KM clustering. Hierarchical K-Means (HKM) clustering is validated by using FOM methodology for assessing the quality of clusters. As this clustering allows clustering the genes correspond to their expression levels and no details about the conditions or samples which may also the factors affecting the expression level of genes.

For simultaneous clustering of genes and conditions the biclustering is performed for same gene expression data. Again biclustering is performed using Hierarchical-K-Means (HKM). Here the sub-matrices of genes and columns are found. The data matrix, A, with set of rows X and set of columns Y, where the elements $a_{ij}$ corresponds to a value representing the relation between row i and column j. The matrix A, with n rows and m columns, is defined by its set of rows, $X=\{x_{1}...x_{n}\}$ $Y=\{y_{1,...}y_{m}\}$ using (X,Y) to denote the matrix A. If $I \subseteq X$ and $J \subseteq Y$ are subsets of columns, respectively, $A_{IJ}=(I,J)$ denotes the submatrix $A_{IJ}$ of A that contains only the elements $a_{ij}$ belonging to the sub-matrix with set of rows I and set of columns J.

The data matrix A a cluster of rows is a subset of rows that exhibit similar behavior across the set of all columns. This means that a row cluster I= $\{i_1…i_k\}$ is a subset of rows ( $I \subseteq X$ and $k \leq n$ ).A cluster of rows (I, Y) can thus be defined as a k by m sub-matrix of the data matrix A. Similarly, a cluster of columns is a subset of columns that exhibit similar behavior across the set of all rows. A cluster $A_{XJ}=(X, J)$ is a subset of columns defined over the set of all rows X, where J= $\{j1_{,...}j_s\}$ is a subset of columns ( $J \subseteq Y$ and $s \leq m$ ).A cluster of columns (X,J) can be defined as an n by s sub-matrix of the data matrix A.

A HKM bicluster is a subset of rows that exhibit similar behavior across a subset of columns, and vice-versa. The HKM bicluster $A_{IJ}= (I, J)$ is a subset of rows and a subset of columns where I= $\{i_1…i_k\}$ is a subset of rows ( $I \subseteq X$ and $k \leq n$ ), and J=$\{j_{1,...}j_s\}$ is a subset of columns ( $J \subseteq Y$ and $s \leq m$ ).A HKM bicluster (I,J) can then be defined as a k by s sub-matrix of the data matrix A.A set of HKM biclusters $B_k=( I_k, J_k)$ is identified such that each HKM bicluster $B_k$ satisfies some specific characteristics of homogeneity

For discovering the knowledge the BLAST tool is utilized. The subsequences in the database, which are similar to the query, are found. The main idea of BLAST is that there are often high-scoring segment pairs (HSP) contained in a statistically significant alignment. It searches for high scoring sequences alignments between the query sequences using a heuristic approach. The BLAST program uses a heuristic approach that is less accurate. The algorithm is as follows:

1. Remove low-complexity region or sequence repeats in the query sequence.
2. In the query sequence make the list of possible matching genes.
3. Organize the remaining high-scoring genes into an efficient search tree.
4. Repeat steps for each gene in the query sequence.
5. Scan the database sequences for exact match with the remaining high-scoring genes.
6. Extend the exact matches to high- scoring segment pair (HSP).
7. List all the HSP's in the database whose score is high enough to be considered.
8. Evaluate the significance of the HSP score.
9. Make two or more HSP regions into a longer alignment.
10. Show the gapped local alignments of the query and each of the matched database sequences.
11. Report the matches whose expect score is lower than a threshold parameter E.

Thus, the complete knowledge about genes from input gene expression data is obtained by performing biclustering and clustering.

## 5 ALGORITHMS

In proposed work algorithm, the HKM clustering algorithm is performed and also biclustering is performed for HKM clustering algorithm. The obtained clusters are validated and the knowledge discovery is done. The algorithm for proposed work follows:

**HKM Biclustering Algorithm**
**Input:** Gene Expression data;
**Output:** Informative (knowledge) clusters;
Begin
Similarity matrix calculation using Pearson correlation coefficient;
HC ();
{ /* Hierarchical clustering */Start by assigning each item to a cluster, if N items are there in the input data set, then N clusters will be produced;

Each contains just one item. Let the distances (similarities) between the clusters the same as the distances (similarities) between the items they contain.
1. Find the closest (most similar) pair of clusters and merge them into a single cluster (using single linkage clustering).
2. Compute distances (similarities) between the new cluster and each of the old clusters.
3. Repeat steps 2 and 3 until all items are clustered into a single cluster of size N and a tree is produced.
4. Tree nodes are mapped to hash table;
}
KM ();
{
/* K-Means clustering for the hash table data (partial clusters) */
1. Initialize K value from hash table entries. Also, the headers of the hash table refer to the centroids.
2. Assign each object to the group that has the closest centroid.
3. When all objects have been assigned, recalculate the positions of the K centroids.
4. Repeat Steps 2 and 3 until the centroids no longer move. This produces a separation of the objects into groups from which the metric to be minimized can be calculated.
}
Knowledge discovery ( );
{
Sequence similarity search for the obtained clusters;
Display knowledge;
}
End.

## 6 RESULTS AND DISCUSSION

The HKM algorithm performs clustering and biclustering of gene expression data. This algorithm is applied to the Yeast cell cycle dataset shows the fluctuation of expression levels of approximately 317 genes over two cell cycles(17 times points) ranges from 10 min to 24 hours. At first the HKM clustering is performed.

The Pearson's Correlation Coefficient distance measurement is chosen to measure the distance between DNA genes and complete runs of the datasets are formed. The clusters formed at the end of HKM clustering are shown in the table 1.

**Table 1
HKM clustering**

| Genes | Cluster Number |
|---|---|
| 1,2,3,4,5,6,7,8 | 1 |
| 9,10,11,12,13,14 | 2 |
| 15,16,17,18,19,20,2 | 3 |

The above HKM clusters formed by HKM clustering algorithm on this dataset are validated by FOM methodology. The validation result shows a steep decline of FOM's for the clusters in graph as shown below. This HKM algorithm achieved the lowest FOM's on this data. The steep decline is indicated in the Fig. 1. Hence it is proved that this algorithm with lower FOM produces high quality clusters.

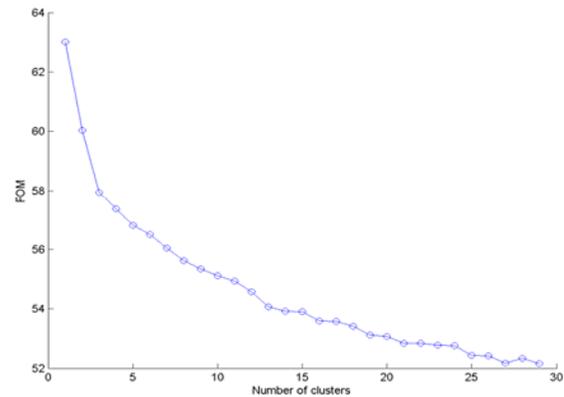

**Fig. 1. FMO's of HKM clustering algorithm on yeast cell cycle dataset**

Then the biclustering is performed for HKM algorithm. As HKM clusters are validated, the HKM biclustering results are also validated. This algorithm also shows a steep decline of FOM's for the clusters in graph as shown below in the Fig. 2.

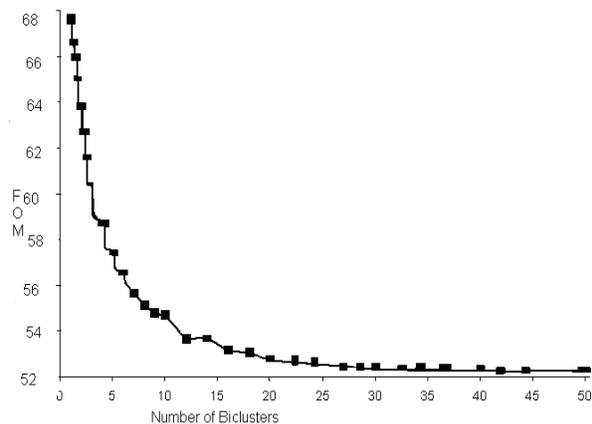

**Fig. 2. FMO's of HKM biclustering algorithm on yeast cell cycle dataset**

After the HKM and HKM biclustering performance, the similarity searching for protein sequence and amino acid sequence for the genes are obtained by using BLAST. The limitation of the proposed approach is that validation metric used here gives the related information in the conditions used to produce clusters. In some situations it is not applicable because, if all experimental conditions contain independent information then the predictive approach is not possible.

## 7 CONCLUSION

In this work, a quality data driven framework for clustering gene expression data is developed. The clustering and biclustering algorithm are validated by FOM methodology. The quality of clusters is evaluated and the result shows relatively high quality clusters. The global model and local model of genes can be obtained by performing both clustering and biclustering. Moreover, the biclustering is outperforming in detecting clusters with higher biological significances than the HKM clustering. It discovers the complete knowledge about the genes from the input gene expression data. The empirical results showthat clustering and biclustering for the same data set yield the complete functional organization of genes and their biological significance.

In future this work can be extended to real gene expression datasets for exploring new biological processes. Further, soft computing techniques, parallel genetic algorithm and classification techniques can be combined to obtain robust clustering and accurate clustering result.

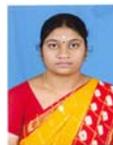

**H.Swathi** received the B.E degree in Electrical and Electronics Engineering from Anna University, Chennai in the year of 2007. She received the M.E degree in Software Engineering from Anna University, Coimbatore in the year of 2009. She is currently working as Lecturer in the Department of Information Technology, Vivekanandha College of Engineering for Women, Tiruchengode. Her current research interests include various aspects of theoretical, methodological and applied research in Data mining, Bioinformatics and Sotware Engineering.